\documentclass[11pt,twoside]{article}

\usepackage{asp2006}
\usepackage{graphicx,natbib}
\bibpunct{(}{)}{;}{a}{}{,}

\begin{document}
\title{Interactions between reversed granulation, $p$-modes, and magnetism?}
\author{Alfred~G.~de~Wijn and Scott~W.~McIntosh}
\affil{High Altitude Observatory, National Center for Atmospheric Research, P.O.~Box 3000, Boulder, CO 80307, USA, dwijn@ucar.edu}
\author{Bart De Pontieu}
\affil{Lockheed Martin Solar and Atmospheric Laboratory, 3251 Hanover Street, Org.~ADBS, Building 252, Palo Alto, CA 94304, USA}

\begin{abstract}
We investigate features that are observed in Ca~\textsc{ii}~H sequences from Hinode in places where reversed granulation seems to interact with $p$-modes.
These features appear ubiquitously in the quiet sun.
They are co-spatial with reversed granulation, and display similar general properties, but have sharper edges and show fast brightness changes.
They also appear predominantly above wide intergranular lanes, indicating a potential connection with magnetism.
We report on the appearance and dynamics of these features using high-resolution, high-cadence observations from Hinode, and we discuss their possible origin.
\end{abstract}

\section{Introduction}
Observations of Ca~\textsc{ii}~H and K line intensity in quiet internetwork areas show a pattern that is roughly the inverse of the underlying granulation called ``reversed granulation''
	\citep[e.g.,][and references therein]{2004A&A...416..333R}.
The anticorrelation is excellent, especially if a 3-minute delay for propagation is taken into account.
Reversed granulation has long been known to be a result of overturning convection
	(\citealp[e.g.,][]{1985SoPh..100..209N,1989A&A...213..371S};
	\citealp[more recently][]{2007A&A...461.1163C}).
Reversed granulation is a hydrodynamic phenomenon that does not require the presence of magnetic field.
Purely hydrodynamic simulations that reproduce reversed granulation well also serve to illustrate this point
	\citep[e.g.,][]{2005A&A...431..687L}.

Reversed granulation is perhaps the most conspicuous feature observed in high-resolution Ca~\textsc{ii}~H and K filtergrams of quiet-sun internetwork.
Its ubiquitous nature makes detection of other features, such as internetwork bright points, difficult.
Here, we report on small-scale structures that at first glance appear similar to reversed granulation in Hinode observations of internetwork quiet sun, but with sharper edges and faster brightness changes.

\section{Observations and Analysis}

We investigate image sequences taken with the Solar Optical Telescope on board the Hinode spacecraft
	\citep{2007SoPh..243....3K,
	2008SoPh..249..167T,
	2008SoPh..249..197S,
	2008SoPh..249..233I,
	2008SoPh..249..221S}
in its blue continuum and Ca~\textsc{ii}~H passbands on March~25, 2007.
The observations studied here have a high cadence of 6.4~s necessary to resolve the fast apparent motions of the reversed granulation.
We apply a high-pass filter separately on each pixel in the Ca~\textsc{ii}~H sequence, keeping only frequencies above 14~mHz.
We also filter the Ca~\textsc{ii}~H cube separately using a ``cone filter'' in $k$-$f$-space to isolate apparent supersonic motions.
We then investigate the blue continuum, the Ca~\textsc{ii}~H, and the two filtered data cubes interactively using the ``cube slicer'' XSlice (available in SolarSoft) that dissects the data in $X$--$Y$, $X$--$t$, and $Y$--$t$ slices.

\begin{figure*}[tbp]
\begin{center}
\includegraphics{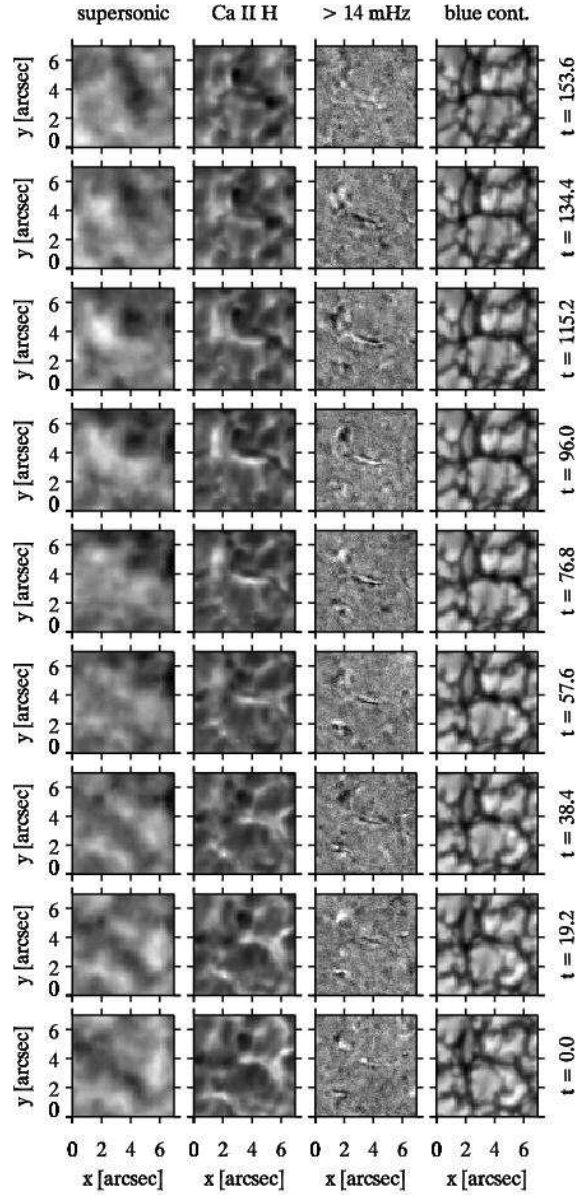}
\caption{Cutouts showing an example ``enhanced reversed granulation'' (ERG).
Columns from left to right: $p$-mode pattern derived from the Ca~\textsc{ii}~H sequence by filtering in $k$-$f$-space for supersonic apparent motions; Ca~\textsc{ii}~H filtergram; Ca~\textsc{ii}~H high-pass filtered pixel-by-pixel; blue continuum filtergram.
Time progresses from top to bottom as indicated on the far right.
The conspicuous brightening in the Ca~\textsc{ii}~H filtergram cutouts is clearly visible in the high-pass filtered panels.
It is located above intergranular lanes.
The $p$-mode pattern shows a maximum at the time of the maximum brightness of the ERG.}
\label{fig:fig1}
\end{center}
\end{figure*}

The filtered data show conspicuous features above intergranular lanes that are typically arc-like in shape with a horizontal extent of several arcsec.
They are co-spatial with reversed granulation, and display similar general properties, but appear to have sharper edges and show fast brightness change.
Their horizontal motions track the underlying intergranular lane closely, with no evident time delay.
We find no obvious recurrent behavior.
An example is shown in Fig.~\ref{fig:fig1}.
We will refer to these features as ``enhanced reversed granulation'' (ERG).

Investigation of the Ca~\textsc{ii}~H sequence suggests that ERG is associated with $p$-mode oscillations.
The emission from the feature is suddenly increased as the $p$-mode maximum moves through it, then returns back to levels similar to the reversed granulation when the $p$-mode moves off.
The leftmost column in Fig.~\ref{fig:fig1} shows the $p$-mode pattern derived by filtering the Ca~\textsc{ii}~H cube in $k$-$f$-space for apparent supersonic motions.
It shows a maximum at the time that the ERG shows its maximum brightness.

ERG seems to happen preferentially over wide intergranular lanes.
The lanes also frequently show broken-up granulation in the vicinity.
Both are typically considered indications of the presence of magnetic field.
While the example in Fig.~\ref{fig:fig1} does not show pronounced broadening of the intergranular lane or broken-up granulation, many others identified in the sequence do show these effects.

\section{Discussion}

The response ERG in the filtered Ca~\textsc{ii}~H data is caused by the relatively sharp edge, horizontal motion, and rapid change in intensity.
The combination of those properties results in step-like change in intensity for each pixel that ensures an excess of power at high temporal frequencies.

We have investigated several similar sequences, and find that they appear ubiquitously in internetwork areas of quiet sun.
None are detected in the network, possibly due to confusion caused by the presence of many fibrils, or possibly because these features do not exist there.

These features may be the result of interactions of $p$-mode oscillations with internetwork magnetic field.
If so, they may form a conduit to channel energy of $p$-mode oscillations into the quiet chromosphere.
It is then tempting to assume that these events correspond to low-lying loops, due to the horizontal extent of the feature.
However, the lack of recurrent behavior suggests that they are not associated with more stable magnetic structures
	\citep{2005A&A...441.1183D}.
Also, an initial investigation of Hinode sequences that include both Ca~\textsc{ii}~H imaging as well as magnetograms did not reveal conspicuous signatures of co-spatial magnetic field.
Photospheric foot points manifested as strong vertical field would be difficult to miss even in magnetograms with low sensitivity.
It is more difficult to rule out a connection with the recently-discovered ``Transient Horizontal Magnetic Fields''
	\citep[THMFs,][]{2008A&A...481L..25I}.
Photospheric THMFs are located inside granules, whereas the features discussed here are always located above intergranular lanes.
However, no THMFs have been detected in the chromosphere, perhaps because of a lack of sensitivity of the measurements, or because they do not exist there.
In any case, there is currently insufficient evidence to make a statement on this point.

We have extracted the $p$-mode pattern from the Ca~\textsc{ii}~H sequence itself.
It would be better to investigate the relation of these features to $p$-mode oscillations using sequences that include Dopplergrams.
Such sequences exist, but suffer from slower cadence that limit their usefulness.
Alternatively, one can attempt to extract the $p$-mode pattern from the photospheric blue-continuum image sequence, in hopes of at least getting an independent measurement, and compare the positions of maximum brightness (i.e., maximum compression) to the locations of ERG.
We will pursue both approaches in the near future.

\acknowledgements
We thank Mats Carlsson for discussions and co-aligned data cubes.
\emph{Hinode} is a Japanese mission developed and launched by ISAS/JAXA, with NAOJ as domestic partner and NASA and STFC (UK) as international partners.
It is operated by these agencies in co-operation with ESA and NSC (Norway).


\begin{thebibliography}{12}
\expandafter\ifx\csname natexlab\endcsname\relax\def\natexlab#1{#1}\fi

\bibitem[{{Cheung} {et~al.}(2007){Cheung}, {Sch{\"u}ssler}, \&
  {Moreno-Insertis}}]{2007A&A...461.1163C}
{Cheung}, M.~C.~M., {Sch{\"u}ssler}, M., \& {Moreno-Insertis}, F. 2007, \aap,
  461, 1163

\bibitem[{{De Wijn} {et~al.}(2005){De Wijn}, {Rutten}, {Haverkamp}, \&
  {S{\"u}tterlin}}]{2005A&A...441.1183D}
{De Wijn}, A.~G., {Rutten}, R.~J., {Haverkamp}, E.~M.~W.~P., \&
  {S{\"u}tterlin}, P. 2005, \aap, 441, 1183

\bibitem[{{Ichimoto} {et~al.}(2008){Ichimoto}, {Lites}, {Elmore}, {Suematsu},
  {Tsuneta}, {Katsukawa}, {Shimizu}, {Shine}, {Tarbell}, {Title}, {Kiyohara},
  {Shinoda}, {Card}, {Lecinski}, {Streander}, {Nakagiri}, {Miyashita},
  {Noguchi}, {Hoffmann}, \& {Cruz}}]{2008SoPh..249..233I}
{Ichimoto}, K., {Lites}, B., {Elmore}, D., {et~al.} 2008, \solphys, 249, 233

\bibitem[{{Ishikawa} {et~al.}(2008){Ishikawa}, {Tsuneta}, {Ichimoto}, {Isobe},
  {Katsukawa}, {Lites}, {Nagata}, {Shimizu}, {Shine}, {Suematsu}, {Tarbell}, \&
  {Title}}]{2008A&A...481L..25I}
{Ishikawa}, R., {Tsuneta}, S., {Ichimoto}, K., {et~al.} 2008, \aap, 481, L25

\bibitem[{{Kosugi} {et~al.}(2007){Kosugi}, {Matsuzaki}, {Sakao}, {Shimizu},
  {Sone}, {Tachikawa}, {Hashimoto}, {Minesugi}, {Ohnishi}, {Yamada}, {Tsuneta},
  {Hara}, {Ichimoto}, {Suematsu}, {Shimojo}, {Watanabe}, {Shimada}, {Davis},
  {Hill}, {Owens}, {Title}, {Culhane}, {Harra}, {Doschek}, \&
  {Golub}}]{2007SoPh..243....3K}
{Kosugi}, T., {Matsuzaki}, K., {Sakao}, T., {et~al.} 2007, \solphys, 243, 3

\bibitem[{{Leenaarts} \& {Wedemeyer-B{\"o}hm}(2005)}]{2005A&A...431..687L}
{Leenaarts}, J. \& {Wedemeyer-B{\"o}hm}, S. 2005, \aap, 431, 687

\bibitem[{{Nordlund}(1985)}]{1985SoPh..100..209N}
{Nordlund}, A. 1985, \solphys, 100, 209

\bibitem[{{Rutten} {et~al.}(2004){Rutten}, {De Wijn}, \&
  {S{\"u}tterlin}}]{2004A&A...416..333R}
{Rutten}, R.~J., {De Wijn}, A.~G., \& {S{\"u}tterlin}, P. 2004, \aap, 416, 333

\bibitem[{{Shimizu} {et~al.}(2008){Shimizu}, {Nagata}, {Tsuneta}, {Tarbell},
  {Edwards}, {Shine}, {Hoffmann}, {Thomas}, {Sour}, {Rehse}, {Ito},
  {Kashiwagi}, {Tabata}, {Kodeki}, {Nagase}, {Matsuzaki}, {Kobayashi},
  {Ichimoto}, \& {Suematsu}}]{2008SoPh..249..221S}
{Shimizu}, T., {Nagata}, S., {Tsuneta}, S., {et~al.} 2008, \solphys, 249, 221

\bibitem[{{Steffen} {et~al.}(1989){Steffen}, {Ludwig}, \&
  {Kruess}}]{1989A&A...213..371S}
{Steffen}, M., {Ludwig}, H.-G., \& {Kruess}, A. 1989, \aap, 213, 371

\bibitem[{{Suematsu} {et~al.}(2008){Suematsu}, {Tsuneta}, {Ichimoto},
  {Shimizu}, {Otsubo}, {Katsukawa}, {Nakagiri}, {Noguchi}, {Tamura}, {Kato},
  {Hara}, {Kubo}, {Mikami}, {Saito}, {Matsushita}, {Kawaguchi}, {Nakaoji},
  {Nagae}, {Shimada}, {Takeyama}, \& {Yamamuro}}]{2008SoPh..249..197S}
{Suematsu}, Y., {Tsuneta}, S., {Ichimoto}, K., {et~al.} 2008, \solphys, 249,
  197

\bibitem[{{Tsuneta} {et~al.}(2008){Tsuneta}, {Ichimoto}, {Katsukawa}, {Nagata},
  {Otsubo}, {Shimizu}, {Suematsu}, {Nakagiri}, {Noguchi}, {Tarbell}, {Title},
  {Shine}, {Rosenberg}, {Hoffmann}, {Jurcevich}, {Kushner}, {Levay}, {Lites},
  {Elmore}, {Matsushita}, {Kawaguchi}, {Saito}, {Mikami}, {Hill}, \&
  {Owens}}]{2008SoPh..249..167T}
{Tsuneta}, S., {Ichimoto}, K., {Katsukawa}, Y., {et~al.} 2008, \solphys, 249,
  167

\end{thebibliography}
\end{document}